\documentstyle[11pt,newpasp,twoside,psfig]{article}
\index{summary}
\index{instructions}
\index{template}

\def\edcomment#1{\iffalse\marginpar{\raggedright\sl#1\/}\else\relax\fi}
\reversemarginpar

\begin{document}
\title{A 2.4\,--\,80\,$\mu$m spectrophotometric study of SS\,433 with ISO}
 \author{Ya\"el Fuchs, Lydie Koch-Miramond}
\affil{Service d'Astrophysique CEA/Saclay, Orme des Merisiers B\^at. 709, 91191 Gif-sur-Yvette cedex, France}
\author{P\'eter \'Abrah\'am}
\affil{Konkoly Observatory, \mbox{P.O. box 67,} 1525 Budapest, Hungary\\
Max-Planck-Institut f\"ur Astronomie, 69117 Heidelberg, Germany}

\begin{abstract}
	We present ISOPHOT spectrophotometric observations of SS\,433 at
	four different orbital phases in 1996 and 1997. The
	He{\footnotesize I}\,+\,He{\footnotesize II} lines\ \ in\ \ both\ \ 
	spectra\ \  of\ \ SS\,433\ \ and\ \ the\ \  Wolf-Rayet\ \ star\ \ WR\,147,\ \ a \mbox{WN8+BO5}
	binary system, closely match. The 2.5\,--\,12\,$\mu$m continuum
	radiation is due to an expanding wind free-free emission in an
	intermediate case between optically thick and optically thin
	regime. A rough mass loss rate evaluation gives about $\sim
	1.4 \times 10^{-4}\,\mathrm{M}_\odot.\mathrm{yr}^{-1}$.
	Results are consistent with a Wolf-Rayet-like companion to the
	compact object in SS\,433.
\end{abstract}
\vspace*{-0.4cm}
\section{Introduction: SS\,433 optical and near-infrared stationary spectra}
	The X-ray binary source SS\,433 and its relativistic jets have
	been studied at many wavelengths, with yet no consensus about
	the component masses and the evolutionary status of the system
	(see Margon 1984 for a review). As shown by van den Heuvel,
	Ostriker, \& Petterson\,(1980) the optical stationary spectrum
	and photometric characteristics of SS\,433 are consistent with
	those of an Of or Wolf-Rayet (WR)-like star with an extremely
	large rate of stellar wind mass loss. The predicted mass
	transfer rates are much larger than the likely mass loss rates
	in the precessing jets (Begelman et al. 1980). King, Taam, \&
	Begelman\,(2000) argue that most of the transferred mass is
	lost from the accretion flow at large radii and is presumably
	the source of the stationary H$\alpha$ line and the associated
	free-free continuum in the near-infrared seen by Giles et
	al.\,(1979).
	
     We\ \ present\ \ middle\ \ and\ \ far-infrared\ \ spectrophotometric\ \ 
	observations\ \ of\ \ SS\,433 with ISOPHOT (Kessler et al. 1996;
	Lemke et al. 1996) to test this interpretation. We will
	discuss the spectral shape of the continuum and compare the
	emission lines in SS\,433 and the binary Wolf-Rayet star
	WR\,147.

\section{Observations}
\subsection{Spectrometry and photometry with ISOPHOT (2.4\,--\,80 $\mu$m)}
	We took ISOPHOT archived data on SS\,433 which was
	observed by ISO on 1996 November 3; 1997 April 11, 17, 23 each
	time during 2156\,s total; the binary orbital phases were
	respectively 0.11, 0.27, 0.71, 0.16 with uncertainty 0.19
	according to the ephemeris of Kemp et al.\,(1986) (phase 0
	corresponds to the eclipse of the accretion disk by the normal
	star) ; the precession phases were respectively 0.84, 0.81,
	0.85, 0.89 (phase 0 when the approching jet lies farthest to
	the line of sight). The observing modes were ISOPHOT-S
	spectrophotometry in the ranges 2.4\,--\,4.8 and 6\,--\,12
	$\mu$m with spectral resolutions $\sim$ 0.04 and 0.1 $\mu$m
	respectively, and ISOPHOT multi-filter photometry at 12, 25,
	60 $\mu$m.  The four 2.4\,--\,12 $\mu$m SS\,433 spectra are
	shown in Fig.\,1a; they are de-reddened according to Lutz et
	al.\,(1996) assuming Av = 8 (Margon 1984).
%	The equivalent widths of the emission lines are given on Table\,1.\\
\vspace*{-0.3cm}
\begin{figure}[!htp]
%\centerline{\psfig{figure=aguerosm_1.ps,height=4.0in}}
\centerline{\hbox{\hspace{0cm}\psfig{figure=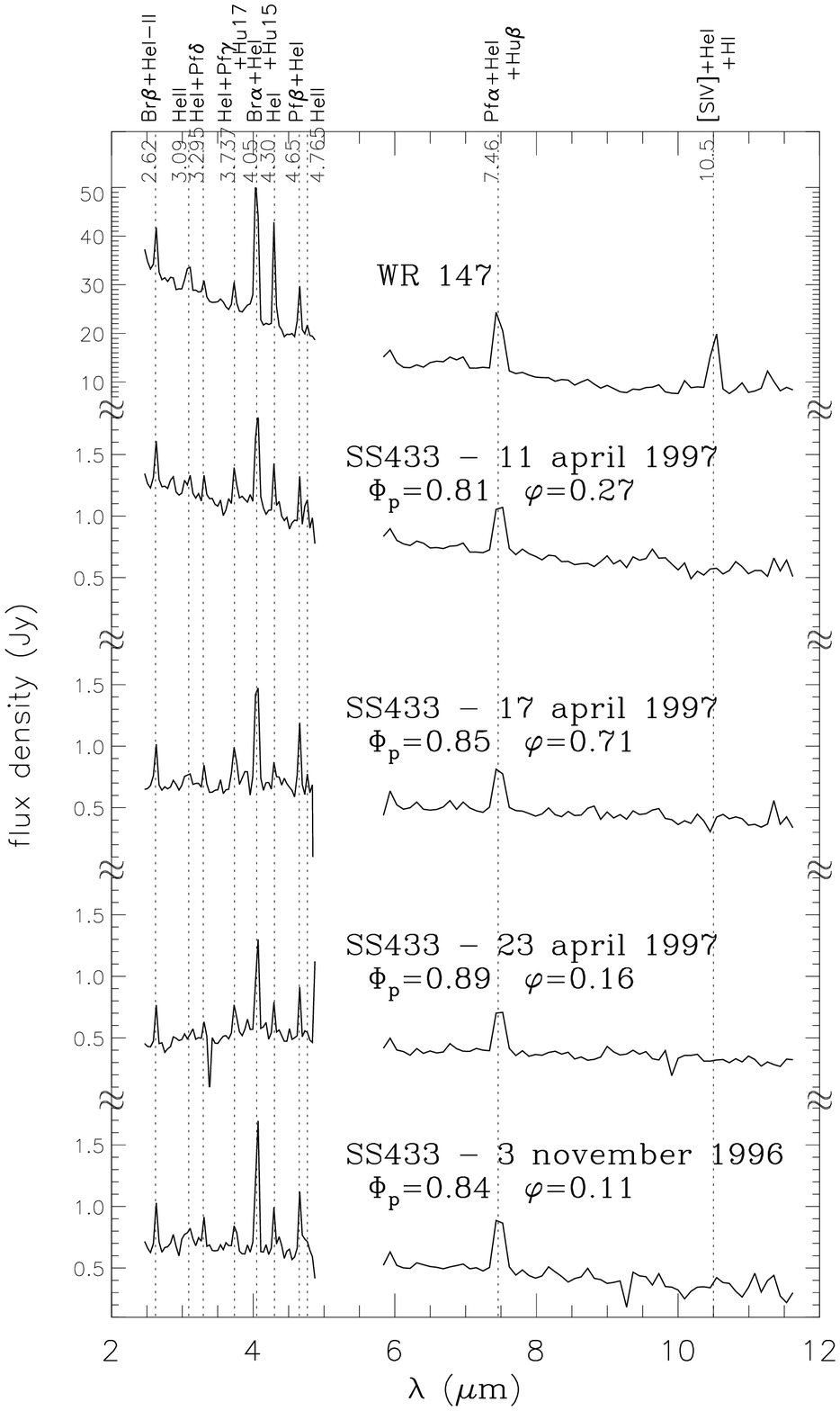,width=6.2cm}\hspace{0cm}\psfig{figure=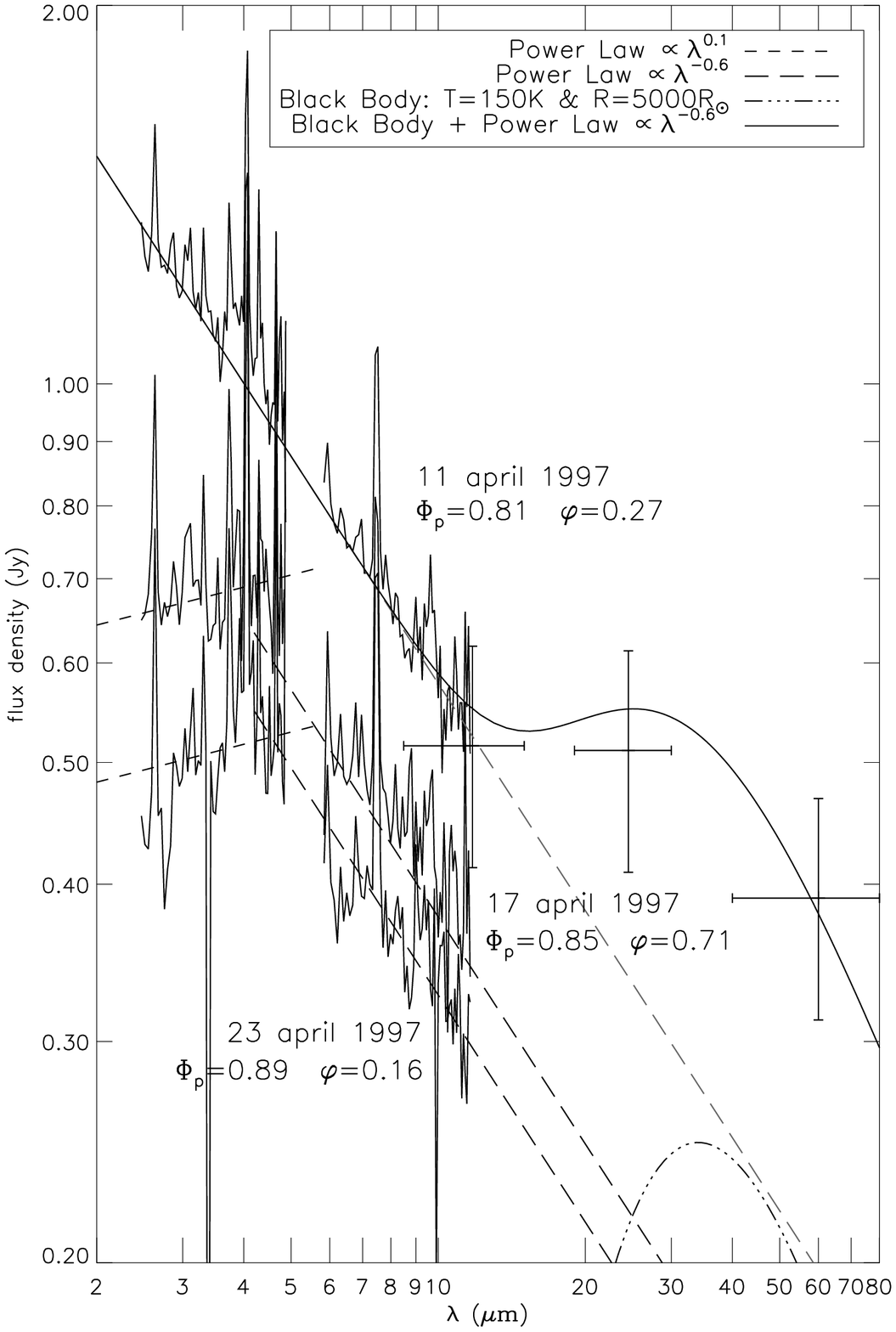,width=6.5cm}}}
\vspace*{-0.2cm}
\caption{a (left): comparison between WR147 and SS\,433 spectra. b (right): simple modeling of April 1997 SS\,433 spectra (see text).}
\label{fig1}
\end{figure}

\vspace*{-0.5cm}
\subsection{Comparison with the WR\,147 spectrum}
   	Several Wolf-Rayet stars were observed with the high spectral 
   	resolution Short Wavelength Spectrometer (SWS) on ISO, which
   	archival data were smoothed %(using an IDL routine of B. Schulz from VILSPA) 
	to the lower resolution of
   	ISOPHOT-S and compared to SS\,433 spectra. The lines detected
   	in these latter spectra closely match the spectrum of the Wolf-Rayet
   	WR\,147, a WN8+B05 binary system (see Fig.\,1a), except for
   	the \mbox{10.5\,$\mu$m} [S{\footnotesize IV}]+He{\footnotesize
   	I}+H{\footnotesize I} line missing in SS\,433. The main
   	detected lines, as at 2.62, 4.05, 4.30 and \mbox{7.46 $\mu$m} are
   	identified according to Morris et al.\,(2000).
%{\bf equivalent width ?}
%	Their
%        relative equivalent widths can differ by a factor 1.3 to 3
%        between the SS\,433 phases observed and by a factor 1.3 to 5
%        with the WR star, as shown on Table\,1.\\
\vspace*{-0.1cm}
\section{Interpretation}
\subsection{Spectral fitting of the mid-infrared continuum flux of SS\,433}
        Fitting of the continuum flux of SS\,433 on 1997 April 11, 17,
        23, is shown on Fig.\,1b (the 1996 spectrum is very similar to
        the 17 April 1997 one). In the 4\,--\,12\,$\mu$m range, the
        slope ($\alpha$ with F$_\nu \sim \nu^\alpha$) is in excellent
        agreement with the 0.6 value expected from the free-free
        emission of an extended optically thick envelope possibly far
        from spherical (Schmid-Burgk 1982); between 2.4 and 4 $\mu$m
        on 17 and 23 April, the slope becomes flatter $\alpha \simeq
        -0.1$ as going to optically thin free-free emission ; above 12
        $\mu$m, observed only with broad-band filters on April 11, a
        black-body from dust at T = 150\,K (R = 5000\,R$_\odot$) has
        been added to the optically thick free-free emission, although
        this far-infrared emission is also consistent with optically
        thin dust at 120\,--\,250\,K.

\subsection{Rough mass loss rate evaluation}
	As Ogley, Bell-Burnell, \& Fender\,(2001) for Cyg X-3, we evaluated the mass loss
	rate of this free-free emitting wind in SS\,433, following the
	Wright \& Barlow\,(1975) formula (8). With a distance D =
	3.2\,kpc, a Gaunt factor g $\sim$ 1, a \mbox{F$_\nu$ = 1000\,mJy}
	flux at 4\,$\mu$m ($7.5 \times 10^{13}$\,Hz), and for a
	WN-type wind (where the mean atomic weight per nucleon $\mu =
	1.5$, the number of free electrons per nucleon $\gamma_e = 1$
	and the mean ionic charge Z = 1) with a velocity of $v_\infty
	= 1500$\,km.s$^{-1}$,  we find :
%\begin{center}
$\dot {\mathrm M} = 1.4 \times 10^{-4} \ {\mathrm M}_{\odot}.{\mathrm yr}^{-1}.$
%\end{center}
	This is in good agreement with the mass transfer rate
	estimated by van den Heuvel et al.\,(1980) assuming a normal
	homogeneous WR wind, or with recent mass transfer values
	obtained from simulations of SS\,433 evolution by King et
	al.\,(2000). However, this result is higher than the recent
	revised WN mass-loss rate estimates, which have been lowered
	by a factor 2 or 3 due to clumping in the wind (Morris et
	al. 1999).

\subsection{Radio to optical spectrum of SS\,433}
	 Radio to optical continuum spectrum of SS\,433, with data taken
	 at very different epochs, is shown on Fig.\,2 where the
	 source variability is clearly seen. Note that the
	 lower fluxes at 1.6 and 5 GHz of Paragi et al.\,(1999)
	 measured with the VLBA + VLA are explained by the highest
	 angular resolution spectral mapping of the inner core and
	 jets of the binary system. The near-infrared and optical
	 wavelengths slopes are steeper ($\alpha \sim$ 1\,--\,3) than the
	 one observed in mid-infrared with our ISOPHOT spectra. Giles
	 et al.\,(1979) and McAlary \& McLaren\,(1980) interpreted these
	 B to K band filter fluxes as the addition of a hot black body
	 (a 15\,700\,K star or the accretion disk ?) and an optically
	 thin free-free emission.

\section{Conclusions}
	We have shown that the mid-infrared continuum of SS\,433 between
	2.5 - 12 $\mu$m can be explained by the free-free emission of
	an expanding wind in the intermediate case between optically
	thick and optically thin regime. This result is consistent
	with the assumed large mass expelled from the super-Eddington
	accretion disk at large radii (King et al. 2000), and with the
	wind-like equatorial outflow observed by Paragi et al.\,(1999)
	- and their origin is probably a stellar wind from the
	companion star. The close match between the HeI + HeII
	emission lines detected in SS\,433 and WR\,147 is consistent with
	a Wolf-Rayet-like companion to the compact object. Thus SS\,433
	might be the second known X-ray binary containing a 
	Wolf-Rayet star with a compact object after the galactic
	relativistic jet source Cygnus X-3 (Hanson, Still, \& Fender 2000 and
	references therein).
%\section{Examples}
%\acknowledgments
%y'a besoin ?
\vspace*{-0.3cm}
\begin{figure}[!htp]
%\centerline{\psfig{figure=aguerosm_1.ps,height=4.0in}}
\centerline{\psfig{figure=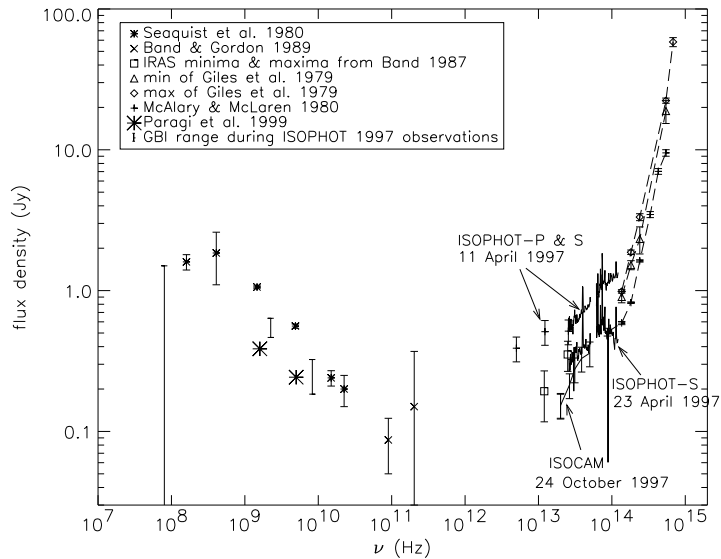,angle=-90,width=10cm}}
\vspace*{-0.4cm}
\caption{Radio to optical spectrum of SS\,433}
\label{fig2}
\end{figure}

\end{document}